\documentclass[preprint,authoryear]{elsarticle}

\usepackage{textcomp}
\usepackage{soul}
\usepackage{comment}
\usepackage{amssymb,amsmath,amsfonts,bm,mathrsfs}
\usepackage{graphicx,stfloats}
\graphicspath{{figs/}}
\usepackage{float}

\usepackage{cite,url,enumerate,xspace}
\usepackage{array,multirow,dcolumn,booktabs,longtable,xcolor}
\usepackage{natbib}

\usepackage[hmargin=1.25in, vmargin=1in]{geometry}
\usepackage{lineno}
\usepackage{setspace}
\newcommand{\PreserveBackslash}[1]{\let\temp=\\#1\let\\=\temp}
\newcolumntype{C}[1]{>{\PreserveBackslash\centering}m{#1}}
\newcolumntype{d}[1]{D{.}{.}{#1}}

\newcommand{\x}{\+x}

\newcommand{\E}[2][]{\ensuremath{%
  \if\relax#1\relax\mathbb{E}\!\left[#2\right]
  \else\mathbb{E}_{#1}\left[#2\right]\fi}}

\newcommand{\equationref}[1]{equation~\eqref{#1}}
\newcommand{\figref}[1]{Figure~\ref{#1}}
\newcommand{\Figref}[1]{Figure~\ref{#1}}

\newcommand{\secref}[1]{Section~\ref{#1}}
\newcommand{\Secref}[1]{Section~\ref{#1}}

\newcommand{\incidents}{\textbf{e}\xspace}
\newcommand{\sacts}{\ensuremath{\mathbf{a}}\xspace}
\newcommand{\safeacts}{\ensuremath{\mathbf{a^+}}\xspace}
\newcommand{\unsafeacts}{\ensuremath{\mathbf{a^-}}\xspace}

\newcommand{\unsafeobs}{\ensuremath{\mathbf{a^-_\mathit{X}}}}

\newcommand{\baseRate}{\ensuremath{\lambda_{*}}}
\newcommand{\vulnweight}[2]{$w_\text{#1}=#2$}

\DeclareMathAlphabet{\mymathbb}{U}{BOONDOX-ds}{m}{n}
\newcommand{\ones}{\mymathbb{1}}

\makeatletter
\def\fps@figure{tbp}
\def\fps@table{tbp}
\makeatother

\journal{Safety Science}

\begin{document}

\begin{frontmatter}
\title{Methodology for Testing and Evaluation of Safety Analytics Approaches}
\author{Antonio~R.~Paiva}\ead{antonio.paiva@exxonmobil.com}
\author{Ashutosh~Tewari\corref{cor1}}\ead{ashutosh.tewari@exxonmobil.com}
 \cortext[cor1]{Corresponding author}
 \address{Corporate Strategic Research, ExxonMobil Research and Engineering Company,\\
   1545 Route 22 East, Annandale, NJ 08801, USA}
 

\begin{abstract}
 There has been a significant increase in the development of data-driven safety analytics (SA) approaches in recent years. In light of these advances it has become imperative to evaluate such approaches in a principled way to determine their merits and limitations. To that end, we propose an evaluation methodology underpinned by a simulated environment that allows for a comprehensive assessment of SA approaches. While assessing such approaches with historical field data is undoubtedly important, such an assessment has limited statistical power because it corresponds to only a few realizations. The proposed methodology enables validation over a large number of realizations, thereby circumventing the statistical limitations of evaluation on historical data. Moreover, a simulated environment allows for a comparison under controlled circumstances, resulting in a fair and systematic assessment of the potential long-term benefits of SA approaches. We demonstrate the utility of the proposed methodology via a case study that compares a few candidate SA approaches, which differ in how they assimilate field  data to assess safety risk and suggest mitigative actions. We show that the proposed methodology indeed reveals useful insights and quantifies the relative merits and drawbacks of the different SA approaches, which would be otherwise difficult to objectively determine in a real-world scenario. 
\end{abstract}


\end{frontmatter}



\section{Introduction}\label{sec:intro}

A crucial aspect of any safety management system~(SMS) is the continuous assessment of work conditions and safety behaviors to identify potential risks and to develop strategies to address them~\citep{li2018safety}. There are a number of strategies by which organizations seek to identify risks such as identifying potential hazards, widespread reporting of incidents and near-misses,  establishing best practices and guidelines for performing certain types of work etc. These strategies are crucial to encourage a ``safety-first" mindset and provide targeted feedback that helps to identify safety areas with higher risk or best practices needing revision. In spite of these efforts, many organizations have been struggling with a plateauing trend in their safety performance, as indicated by different metrics, while still experiencing major incidents~\citep{dekker2016examining}. This has motivated recent work on new approaches to better utilize safety data.

In recent years there has been a significant effort to leverage data analytics to assist in decision-making that directly improves safety performance~\citep{huang2018big, wang2019using, wang2019demystifying}, in what has been called the era of Safety~4.0~\citep{wang2021safety}. To that end, a number of safety analytics (SA) approaches have been reported in the literature (see, for example, \citet{zhu2021application, sarkar2020predicting, verma2018data, poh2018375, tixier2016application, tixier2017construction, kakhki2019evaluating, cheng2013applying}). These studies compare and extend different machine learning methods that attempt to forecast incidents or their severity in different industrial settings. Such approaches aim to discover data-driven insights to help safety managers to better understand the safety landscape, for instance by identifying worktypes or areas that might have higher risk of injury. And, through these analyses, they hope to assist in determining how to prioritize which areas should be intervened upon first such that the associated risk is mitigated in the most effective and impactful manner. Since the underlying models learn from historical data, they facilitate an objective analysis and  help prevent potential biases in the interpretation of safety incidents and the circumstances in which they occur~\citep{vanderSchaaf2004biases}. For these reasons, safety data analytics  holds significant promise in improving cross-sector safety performance. Nevertheless, how to evaluate the improvement provided by these SA approaches is an aspect that has not been adequately examined in the safety science literature.

An ideal verification \& validation procedure for SA approaches (to assess if they can be trusted for forecasting safety risks) must evaluate their performance while accounting for several challenges specific to SMSs. Work environments are often dynamic and with evolving risks resulting in non-stationary safety data. This dynamic nature is due to a combination of factors such as changing work types and/or workforce, field interventions or even evolving human behavior towards safety practices. Perhaps the most crucial aspect that impedes assessment of SA approaches is that the historical safety data comprises only a few realizations of an inherently stochastic random process. Additionally, the challenges are compounded by the fact that safety incidents (injuries and fatalities) are usually sparse, fortunately ~\citep{sarkar2020predicting}. Therefore, validation on historical data has limited statistical power and makes it difficult to distinguish between the inherent randomness in the data and a more effective SA approach. It is also worth noting that traditional experimental hypothesis testing techniques are impractical in the context of safety systems. Techniques such as A/B testing, randomized clinical trials, and other forms of two-sample hypothesis testing have long been considered as the ``gold standard'' for evaluating whether a method performs better than another~\citep{rice2006}. While these techniques are valid (in principle) for evaluating SA approaches, it seems irresponsible to expose workers and expensive infrastructure to higher risks due a potentially inferior approach. Most importantly, given the sparseness of safety events, it would take far too long to evaluate an approach to an acceptable level of statistical confidence. For these reasons, a new testing and evaluation methodology is needed to systematically evaluate the merits of different SA approaches, which is the primary contribution of this work. 

To circumvent the aforementioned challenges, we propose a simulation-based methodology to evaluate SA approaches. Central to our proposal is a \emph{safety environment simulator} that emulates various characteristics of a SMS of a typical industrial work environment. Such characteristics include the number of safety areas under consideration, the types and the frequencies of safety observations and the distribution of frequency and severity of incidents. These characteristics are controlled by user-chosen input parameters. Since the underlying characteristics of the simulated environment are determined beforehand, the proposed methodology provides a repeatable setting with a well-defined ground-truth with which to evaluate the effectiveness of different SA approaches. Furthermore, using such a simulated environment, one is able to evaluate a large number of scenarios and assess how different approaches perform in a wide range of work conditions. While using a simulator of a work environment and its SMS is undoubtedly a simplification of the reality, it nevertheless allows for a consistent and systematic comparison of SA approaches under exactly the same conditions which would be otherwise very difficult, if not impossible, to do in the real world. 

The remainder of this paper is organized as follows. First, \secref{sec:background} discusses several general characteristics of SMSs and how they are considered in the context of the methodology proposed herein. The simulated environment and its stochastic formulation are then described in \secref{sec:simulator}.  \Secref{sec:results} presents a case study that demonstrates how  different SA approaches can be objectively assessed using the proposed methodology. The discussion, conclusions, and considerations for future work are given in \secref{sec:discussion}.

\section{Work environment and safety system characteristics}\label{sec:background}

Work environments and their SMSs can differ in many ways. Nevertheless, there are a number of characteristics that are widely common to nearly all SMSs. This section details the general characteristics which will be assumed to be present, the terminology used to refer to them, and the role they play in the overall methodology.
It is important to emphasize that many of the aspects mentioned here should not be taken as prescriptive requirements, but rather as the characteristics of a typical work environment considered in this paper. In fact, we encourage tailoring of the simulator to better reflect the circumstances of other work environments being considered. As long as different SA approaches are benchmarked using the same simulator, the comparative analysis thereof will remain consistent.

\subsection{Safety data}\label{sec:observations}

Safety data is already collected in all work environments, although to varying degrees of frequency and detail, and is used by decision makers to understand safety performance and evaluate potential improvements. Similarly, SA approaches use this data to assess safety risk. Therefore, the central goal of the simulator proposed herein is to generate safety data with realistic interdependencies such that these approaches can be appropriately tested.

Most of the safety data collected in SMSs corresponds to events in the form of either \emph{safety observations} or \emph{incident reports}. These have also been referred as proactive and reactive safety data in the literature~\citep{sarkar2020predicting, verma2018data}, respectively. Safety observations are continuously collected and provide a snapshot of work conditions or workers' actions or behaviors while performing their work. Moreover, we consider that both safe (i.e., `positive') and unsafe/at-risk (i.e., `negative') observations are captured~\citep{grant2018back}, and design the simulator for this general case. Since observations are acquired regardless of whether incidents occur or not, they are crucial indicators of an environment's state from a safety standpoint.

In general, multiple types of observations may be collected. As an example, in this paper we will consider that three types of  observations are collected: work-safety observations~(WSOs), safety assessment observations~(SAOs), and best-practices observations~(BPOs). WSOs could correspond to impromptu observations collected by workers capturing both safe and unsafe behaviors, practices, and situations during normal work. For example, a WSO might be submitted by a worker reflecting on how they reacted, or observed a colleague react, upon realizing that they lacked the proper personal protective equipment~(PPE) for a task. WSOs could also be captured for workplace hazards that need to be addressed. SAOs could correspond to more formal observations in the sense that they are collected by other workers or safety professionals \emph{actively engaged} in assessing safety as other perform their tasks. Because of the more deliberate observing nature, one might expect SAOs to be more significant. On the other hand, it could also be argued that the deliberate task of assessing safety can prime observers to look for negative circumstances, thus potentially introducing biases in the data. BPOs will be taken to correspond to safety inspections, and thus also deliberate observations but primarily focused on the critical evaluation of best practices in the execution of well-defined tasks to ensure that the best practices are appropriate and that they are being followed. More important than the specific meaning attributed to these observation types, these are included here to shown how the simulator can be used to test the robustness of SA approaches if there are biases in data collection.

Incident reports are the other crucial component of safety data because they directly inform us about the frequency and severity of safety incidents, both of which are critical to characterize safety risk. Note that here `incidents' refers to both near-misses and actual incidents (i.e., resulting in injuries, property damage, or environmental issues)~\citep{sarkar2020predicting, verma2018data}. This perspective takes both near-misses and actual incidents to be substantive negative safety events which differ only in the severity of the outcome.

\subsection{Incident severity}\label{sec:hurt.level}

Incident severity is the other crucial dimension in assessing overall safety risk~\citep{zhu2021application, sarkar2020predicting, kakhki2019evaluating}. Accordingly, the simulator must also be able to generate severity levels for each incident so that one can evaluate if a SA approach can correctly assess variability in incidents severity of each safety area.

There are a few possibilities for representing an incident's severity. Traditionally, severities of personal injury incidents were quantified according to a treatment-based approach reflecting the type of medical treatment used in response. While this approach is common among many industries, and is often required in regulatory compliance, it emphasizes administrative reporting and incident escalation management procedures. As a result, low-severity injuries tend to be overlooked and this fails to convey a safety culture of caring and improvement~\citep{smith2013hurt}. Moreover, treatment-based metrics are primarily myopic because they focus only on the severity immediately experienced after an incident occurs. Therefore, a treatment-based severity does not include an assessment of potential consequences, which are critical to understand and prevent future incidents.

In this paper, we consider that incident severity is characterized according to a Hurt-based representation, which has been recently proposed and addresses the limitations of a treatment-based approach~\citep{smith2013hurt, etaje2013efficiency}. The Hurt representation provides a methodology to consistently describe the severity of any actual incident and, perhaps more crucially, evaluate the potential severity under less favorable circumstances. We show how both the \emph{actual Hurt level}~(AHL) and \emph{potential Hurt level}~(PHL) severities can be obtained from the simulator. Moreover, while this representation is focused on injuries, it can be extended to represent severity of an incident pertaining to process, property, and environment. This means that SA approaches seeking to integrate learnings from a broad class of incident types~\citep{eggleston2014assessing} can also be tested by the simulation framework proposed here.
The Hurt-based severities are quantified along 6 levels, zero through five. A Hurt level of zero indicates that no injury took place, meaning that it was a near-miss. A Hurt level of one indicates minor injuries needing only first-aid and involved no lost time. Higher Hurt levels correspond to drastically increasing severity of outcomes, with levels 4 and 5 corresponding to one and multiple fatalities respectively~\citep{smith2013hurt}. Note that, by definition, the PHL is always equal or greater than the AHL.

\subsection{Grouping by safety areas}\label{sec:safety.areas}

Safety observations and incidents are typically grouped according to a set of `safety areas'. These areas may correspond, for example, to different work types and/or different hazard categories commonly encountered while performing a task. They are crucial in providing context to the collected safety data, thus enabling decision making to determine focus priorities. Hence, these safety areas correspond to the dimension toward which safety analytics approaches must yield assessments of risk or help steward interventions.

The statistics of safety data grouped by safety areas typically differ significantly in practice. This is expected as the data may correspond, for instance, to projects of different size or work types. Thus, the simulator accounts for such diversity by specifying a number of factors. These factors include how often the tasks or work associated with a safety area are performed, how frequently those tasks are performed unsafely, and the likelihood that unsafe tasks lead to an incident. The construction of the simulator also specifies the interdependencies between these factors in a way that mimics our understanding of how events unfold in a typical work environment.


\section{Environment simulator}\label{sec:simulator}

The task of simulating a work environment and its SMS is nontrivial for a number of reasons. First and foremost, work environments typically have a wide range of tasks being carried out by a diverse workforce on a daily basis, any of which could result in a safety incident.  Secondly, incidents can occur due to a myriad of reasons ranging from equipment/structure failures, to lack of safe guards, or to at-risk behaviors. Naturally, the underlying processes governing these incidents do not adhere to physical laws that can be modeled mathematically, thus making the task of simulating them more challenging. However, considering that safety data comprises discrete events in time, certain features of safety observations and incidents can be described statistically, as will be done here.

\begin{figure}[tb]
  \centering
  \includegraphics[width=0.9\textwidth]{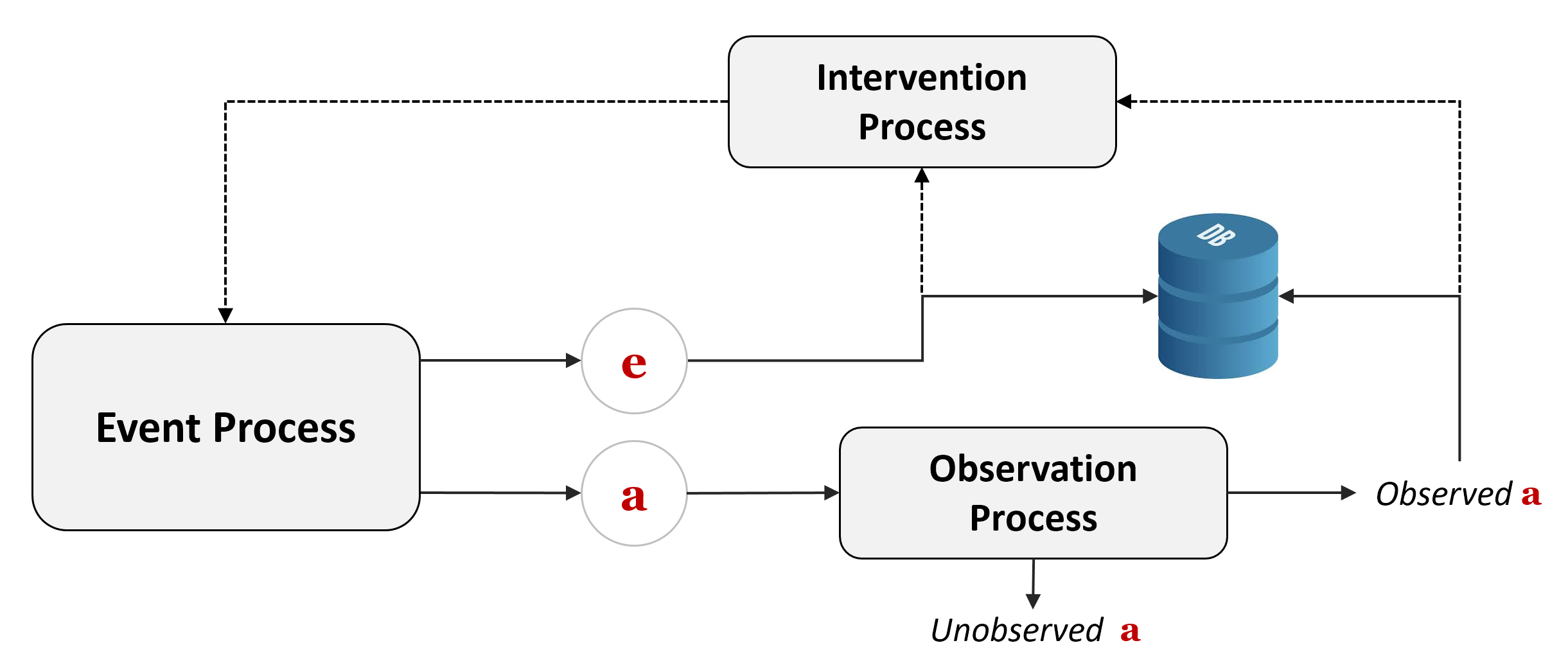}
  \caption{A schematic overview of the proposed environment simulator, which comprises of three primary blocks: the \emph{Event process}, \emph{Observation process} and \emph{Intervention process}. Each of these blocks serves a distinct purpose with a shared goal to make the simulator as realistic as possible.}
  \label{fig:simulator_schematic}
\end{figure}

The simulator will be organized according to the set of safety areas discussed in \secref{sec:safety.areas}. Since safety areas correspond to different tasks, observations and incidents will be characterized and simulated independently between safety areas. This should be an acceptable assumption in most cases. However, if needed, dependencies between safety areas can also be captured by introducing correlations between the temporal dynamics of the simulator's parameters. \Figref{fig:simulator_schematic} provides an overview of the proposed safety environment simulator and the three blocks, named \emph{Event process}, \emph{Observation process} and \emph{Intervention process}, that form its core components. It is worth noting that the statistics for each safety area are controlled by several parameters chosen beforehand. These user-specified parameters have a simple interpretation, and the values assigned to them should ideally be representative of the work environment in consideration. One may also set the simulator parameters to test the robustness of a SA approach under less than ideal or even worst case scenarios.

\subsection{Event Process} \label{subsec:event_process}
We first distinguish three types of events (present in every SMS) that are generated by the event process i.e. \incidents, \safeacts{} and \unsafeacts. The event \incidents{} corresponds to an incident, resulting from a near-miss, an actual injury, or some other loss incident. The event \safeacts{} is a task/situation that is inherently safe because there is no apparent risk. Conversely, \unsafeacts{} is a task that is unsafe or at-risk because it corresponds to the failure to follow safe practices or  someone inadvertently making a mistake~\citep{conklin20195} which could have resulted in an incident. One will generally encounter all of the three types of events for each of the safety areas, albeit at different rates. Note that at this stage the events \safeacts{} and \unsafeacts{} reflect only whether the work/situation itself is safe or unsafe/at-risk. Whether these circumstances are recorded as observations depends on if someone observes the situation and takes the action to record it. This aspect is modeled by the observation process in the next section. By distinguishing between whether the tasks are performed safely (or not) and if they are observed (or not), one can simulate potential biases in data collection~\citep{vanderSchaaf2004biases} and thus test the ability of subsequent SA approaches to cope with them. It is assumed here that events \incidents (actual injuries or near-misses) are always recorded. 

The main purpose of the event process block is to specify the random process that governs the occurrences of events  of the aforementioned types (i.e., \incidents, \safeacts{} and \unsafeacts). In applied statistics, a Poisson process is widely used to describe the occurrence of events distributed over time~\citep{daley1988introduction}. The Poisson process assumes that events are independent from each other, meaning that the occurrence of an event does not impact the probability of occurrence of any other event. Let $\baseRate^i$ be the `task~rate'; that is, the rate at which activities corresponding to the $i^{th}$ safety area are being performed. Then, the Poisson process models the number of occurrences ($n_{task}^i$) of these activities in a time interval $\Delta t$ as Poisson distributed random variables i.e.
\begin{align} \label{Poisson_dist_task}
  P(n_{task}^i = k | \baseRate, \Delta t)= \frac{(\Delta t\baseRate^i)^{k} \exp(-\Delta t\baseRate^i)}{k!}.
\end{align}
In words, this equation yields the probability that $k$ activities occurred in the interval $\Delta t$ (i.e., $n_{task}=k$). Accordingly, the number of activities corresponding to the $i^{th}$ safety area for a given time interval is obtained by sampling from this distribution. The task rate ($\baseRate^i$) can be specified in any unit of time (hour$^{-1}$, day$^{-1}$, year$^{-1}$, etc.), provided that the same unit is used for the time interval $\Delta t$. Note that in our case study we assume $\baseRate^i$ to be a constant. However, it can be a function over time, $\baseRate^i(t)$, because the rate of activities can change over time depending on the stage of a project.

Finally, the different work activities taking place will be assigned to the different event types. This is controlled by the parameters $\xi^i, \alpha^i \in [0, 1]$, where $\xi^i$ is the internal state corresponding to the fraction of activities that are performed unsafely and $\alpha^i$ is the fraction of those that may further lead to actual incidents (\incidents).
The parameter $\alpha_i$ is chosen to be constant here, but in general it could also be a function of time and even include correlations between values across safety areas to make them interdependent.
Using these parameters, the number of incidents ($n_{\incidents}^i$), safe activities ($n_{\safeacts}^i$), and unsafe activities ($n_{\unsafeacts}^i$) can be obtained for each timestep by sampling according to the following generative model,
\begin{align}\label{event_generation_process}
  n_{task}^i &\sim \mathrm{Poisson}(\baseRate^i) \\
  (n_{\incidents}^i,n_{\unsafeacts}^i,n_{\safeacts}^i)|n_{task}^i 
    &\sim \mathrm{Multinomial}\left(n_{task}^i,[\alpha^i\xi^i,\ (1-\alpha^i)\xi^i,\ (1-\xi^i)]\right).
\end{align}
The multinomial distribution yields the counts of the three mutually exclusive events types (\incidents, \unsafeacts, \safeacts) such that the total count adds up to $n_{task}^i$. In fact, by virtue of a standard result from probability theory, each of the counts, $n_{\incidents}^i$, $n_{\unsafeacts}^i$, and $n_{\safeacts}^i$, can also be drawn from separate Poisson distributions as follows
\begin{subequations}
  \begin{align}
    n_{\incidents}^i &\sim \mathrm{Poisson} (\alpha^i \xi^ i \baseRate^i) \label{eq:incident_generation}\\
    n_{\unsafeacts}^i &\sim \mathrm{Poisson}( (1-\alpha^i) \xi^ i \baseRate^i) \\
    n_{\safeacts}^i &\sim \mathrm{Poisson}( (1-\xi^i) \baseRate^i). 
  \end{align}
\end{subequations}
Note that these equations reflect a natural coupling between the three types of events, because the number of events for each type is dependent on the common parameters $\xi^i$ and $\alpha^i$. As one expect, if the proportion of unsafe activities $\xi^i$ increases, the number of incidents \incidents{} will also increase proportionally. Put differently, the above equations are equivalent as if one were to first separate activities into safe and unsafe (according to $\xi^i$), and then determine (according to $\alpha^i$) which unsafe activities lead to incidents \incidents{} and which ones do not (i.e., assigned to \unsafeacts).

\begin{figure}[ht]
  \centering
  \includegraphics[width=0.5\textwidth]{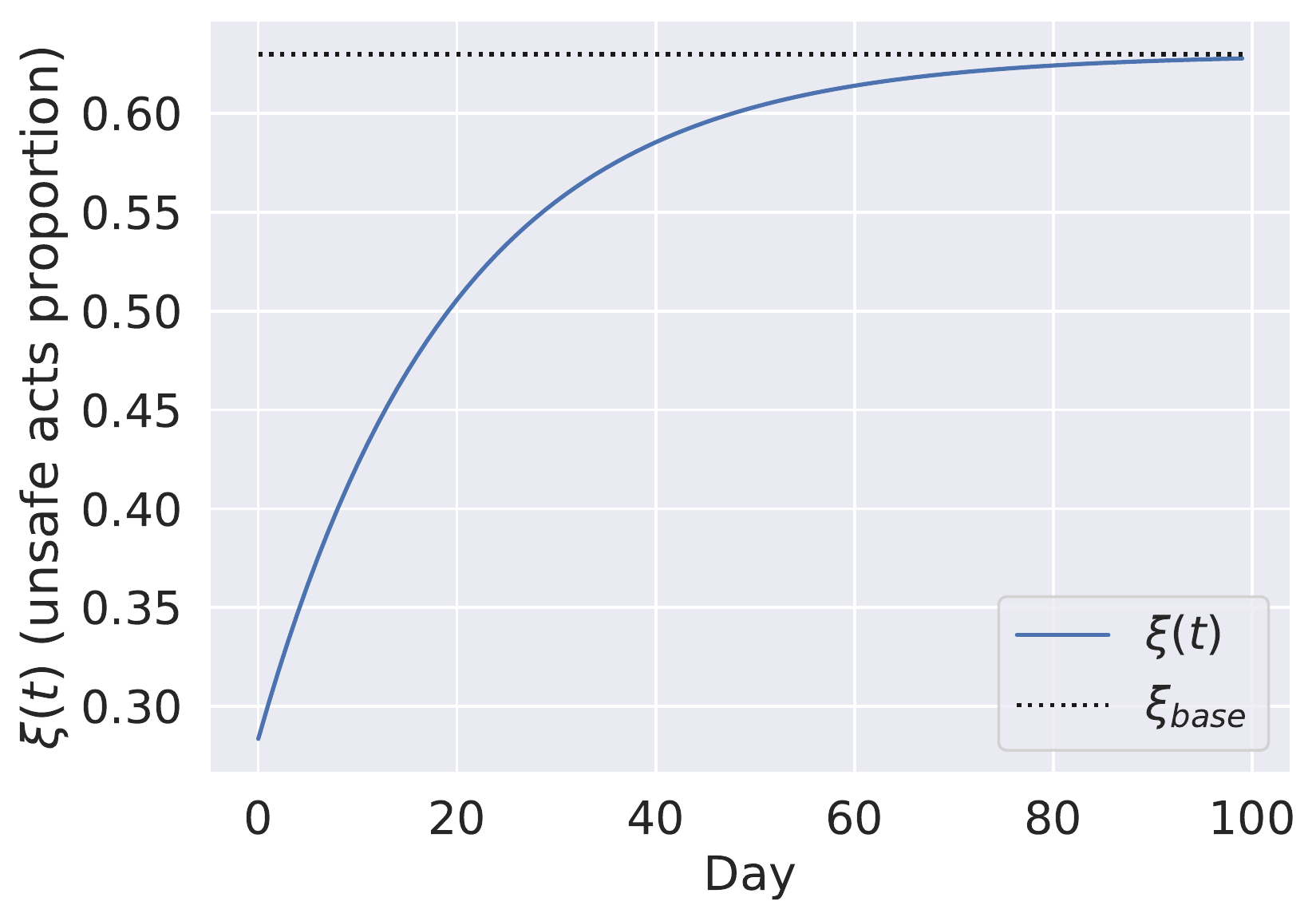}
  \caption{The proposed dynamics of the proportion of tasks being carried out unsafely for an arbitrary safety area. Starting from an initial state ($\xi(0)=0.28$) the state of the safety area continuously worsens (in terms of the proportions of unsafe acts) and converges to the specified worst-case base probability, $\xi_{base}$. This dynamics attempts to capture a typical human trait of becoming progressively careless in the absence of any supervision.}
  \label{fig:xi_dynamics_plot}
\end{figure}

Although in the previous expressions we have suppressed the time dependence of parameter $\xi^i$ (for the ease of exposition), it is in fact a function of time. The time evolution of $\xi^i$  (the probability that an activity is unsafe) can be modeled as,
\begin{align}
  \xi^i(t)&=(1-\theta^i(t))\xi^i_{base} \label{eq:xi_dynamics} \\
  \theta^i(t) &=k^i\theta^i(t-1) \label{eq:theta_dynamics}
\end{align}
where $\xi^i_{base}$ is a given \emph{worst-case} probability that activities are performed unsafely for the $i^{th}$ safety area. In other words, in the absence of any external feedback/intervention, $\xi^i$ will converge to this number. The rationale for having this worst-case probability stems from the expectation that even when completely unmonitored the ratio of unsafe to total activities will stabilize at a certain level. The internal function $\theta^i(t) \in [0,1]$ captures the current ``safety state'' of the $i^{th}$ safety area, with $\theta^i(t)=1$ corresponding to the safest state (i.e., all activities are being safely carried out) and $\theta^i(t)=0$ the most unsafe state (i.e., proportion of unsafe activities is determined by $\xi^i_{base}$). The dynamics of $\xi^i(t)$ are thus controlled through $\theta^i(t)$ by the decay parameter $k^i \in [0,1]$ (cf.\ equations~\eqref{eq:xi_dynamics} and \eqref{eq:theta_dynamics}). This dynamics attempt to mimic a typical human behavior of forgetfulness or a tendency to become complacent over time in the absence of external feedback~\citep{dekker2016drift}. The user-specified parameter $k^i$ determines the rate of such forgetfulness/complacency, with lower values signifying faster convergence of $\theta^i(t)$ toward zero, and $\xi^i(t)$ toward $\xi^i_{base}$ (i.e., faster transition to complacency), and vice versa. For the illustration of this straightforward concept, the time evolution of $\xi$ is plotted  in \figref{fig:xi_dynamics_plot} with the parameter settings of $\xi_{base}=0.63$, $k=0.95$ and $\theta(0)=0.55$ for daily timesteps.

When an incident occurs, a severity level is associated with it. As described in \secref{sec:hurt.level}, a Hurt-representation of severity is used in this paper, which takes values 0 through 5. The severity level describing the incident is referred to as the \emph{actual Hurt level}~(AHL). When an incident occurs the simulator samples a one-hot vector (i.e., one entry is one and all others are zero), where the index of the only non-zero entry indicates the Hurt level. For the $i^{th}$ safety area, this is generated according to a Multinomial distribution over the six Hurt levels,
\begin{align}\label{eq:ahl_generation}
  (\ones^i_{ahl=0}, \ones^i_{ahl=1},\cdots,\ones^i_{ahl=5})  \sim \mathrm{Multinomial}(1, [p^i_{0},p^i_{1},\cdots, p^i_{5}])
\end{align}
where $\ones_{ahl=j}$ denotes the indicator function that yields one if the subscript condition holds and zero otherwise, and $[p^i_{0},p^i_{1}, \cdots, p^i_{5}]$ denote the probability distribution of the 6 Hurt levels for the $i^{th}$ safety area. These probabilities are user-specified to emulate a desired characteristics of severities found in practice. Certain safety areas may involve inherently riskier activities and thus are more likely to result in incidents with higher Hurt levels than others. Such characteristics can be captured in these probability distributions.

Similarly, the simulator also generates the \emph{potential Hurt level}~(PHL) associated with every incident, characterizing the Hurt level that could have possibly resulted under less favorable circumstances. The generative process of the PHL is akin to that of the AHL except that it must ensure that the PHL is greater or equal than the AHL. Hence, one first obtains the AHL value (say $ahl$) and conditions the PHL distribution to enforce the constraint, as shown here,
\begin{align} \label{eq:phl_generation}
   (\ones^i_{phl=0}, \ones^i_{phl=1}, \cdots, \ones^i_{phl=5})| ahl \sim \mathrm{Multinomial}(1, [0, \cdots, 0,\hat{p}^i_{ahl}, \hat{p}^i_{ahl+1}, \cdots, \hat{p}^i_{5}]).
\end{align}
where the Hurt level probability distribution has been modified by zeroing the probability of any Hurt level less than $ahl$ (since $phl\geq ahl$) and renormalizing the remaining values so that they sum to 1.

Finally, while not integral to the generative process, it is convenient to explicitly characterize the distribution of daily incidents per severity. Conditioned on a certain number of incidents, the previous generative process allows us to observe that the conditional distribution of the daily count of incidents for a particular Hurt level $j$, denoted $\mathbb{C}_{ahl=j}^i$, has the form
\begin{align}\label{eq:HL_count_given_incident_count}
  P(\mathbb{C}_{ahl=j}^i \ | \ n_\incidents^i=k) = \mathrm{Binomial}(k, p^i_j).
\end{align}
Then, in order to obtain the marginal distribution of the daily count of incidents of a particular Hurt level, we need to sum over all possible values of incident counts, that is,
\begin{align}
P(\mathbb{C}^i_{ahl=j})  &=  \sum_{k=1}^\infty P(n_\incidents^i=k)P(\mathbb{C}_{ahl=j}^i | n_\incidents^i=k)  \label{eq:marginal_dist_ahlj_count}
\end{align}
where $P(n_\incidents^i)$ is given by \equationref{eq:incident_generation}. In \secref{subsec:health_measures}, we will use this distribution to define various metrics that quantify the safety risks of the simulated environment at any time instant. 

\subsection{Observation Process}\label{subsec:observation_process}

While the previous section described a generative process that emulated the occurrence of activities in a work environment, the observation process attempts to characterize how safety data is recorded by observing that environment. Recall that incidents \incidents{} (near-misses and actual incidents) are assumed to be always recorded. Therefore, the observation process is focused only on events of types \safeacts{} and \unsafeacts. Generally speaking, whether a safe or an unsafe activity (event \safeacts or \unsafeacts) is observed and recorded depends on whether an observer was present at the time and place the activity took place as depicted in \figref{fig:simulator_schematic}. Of course, only a subset of activities are typically observed due to limited resources. To express such resource constraints, we consider that there are $m_X$ observers that can record observations at each timestep (e.g., per day) for each observation type ($X$ being WSO, SAO, or BPO in our case). Moreover, define the observer capacity $\rho_X$ denoting the maximum number of observations one observer can make in that timestep. Also, let $\{s^1, s^2, \cdots , s^n\}$, with $\sum_i^n s_i=1$, be the proportion according to which the observers will be distributed across the $n$ safety areas. This distribution can be used to allocate the observers to safety areas as follows.
\begin{align}
  \left( q^1_X, q^2_X,\cdots, q^n_X \right) \sim \mathrm{Multinomial}(m_X, [s^1, s^2, \cdots, s^n]),
\end{align}
where $q^i_X$ denotes the number of observers that will collect observations of type $X$ for activities of the $i^{th}$ safety area. Hence, out of the total $n^i_\sacts=n_\safeacts^i+n_\unsafeacts^i$ safe and unsafe events that occurred in a given timestep (as determined by event process) $n^i_{\sacts{}_X} = \min(\rho_X q^i_X, n^i_\sacts)$ events will be observed.
Note that the notion of ``assigned observers'' and their ``capacity'' is a modeling representation which could similarly be taken to correspond to an average number of observations per timestep.

To determine which events are ultimately observed, another multinomial distribution can be sampled,
\begin{align}
  \left( \ones^1_X, \ones^2_X, \cdots, \ones^{n^i_a}_X \right)
    \sim \mathrm{Multinomial}(n^i_{\sacts{}_X}, [\mathbf{p}_X^{\safeacts}, \quad \mathbf{p}_X^{\unsafeacts}])
\end{align}
where the $\ones^j_X$ indicator function denotes whether the $j^{th}$ event is recorded by observation type $X$ or not. The positive vectors $\mathbf{p}_X^{\safeacts}$ and $\mathbf{p}_X^{\unsafeacts}$ of length $n_\safeacts^i$ and $n_\unsafeacts^i$, respectively, have components which jointly sum to 1. These probability vectors, in turn, can be sampled from a Dirichlet distribution as follows.
\begin{align}
  [\mathbf{p}_X^{\safeacts}, \quad \mathbf{p}_X^{\unsafeacts}]
    \sim \mathrm{Dirichlet}\left(
      \left[\eta_X^{\safeacts} \times \mathbf{1}^{1\times n_\safeacts^i}, \quad
            \eta_X^{\unsafeacts} \times \mathbf{1}^{1\times n_\unsafeacts^i}
      \right] \right)
\end{align}
The positive real-valued parameters $\eta_X^{\safeacts}$ and $\eta_X^{\unsafeacts}$ are user-specified concentration parameters of the Dirichlet distribution that governs the relative potential bias of favoring recording safe activities instead of unsafe activities, or vice-versa. If $\eta_X^{\safeacts}=\eta_X^{\unsafeacts}$, then the observation type $X$ has no bias. However, if $\eta_X^{\unsafeacts} = 3\times\eta_X^{\safeacts}$ , we are likely to record, on an average, 3 times as many unsafe activities as safe activities for observation type $X$. These parameters can be used to attribute varying levels of biases to different observation types which often exist in real-world SMSs.

\subsection{Intervention Process}\label{subsec:intervention_process}

\begin{figure}[tb]
  \centering
  \includegraphics[width=0.5\textwidth]{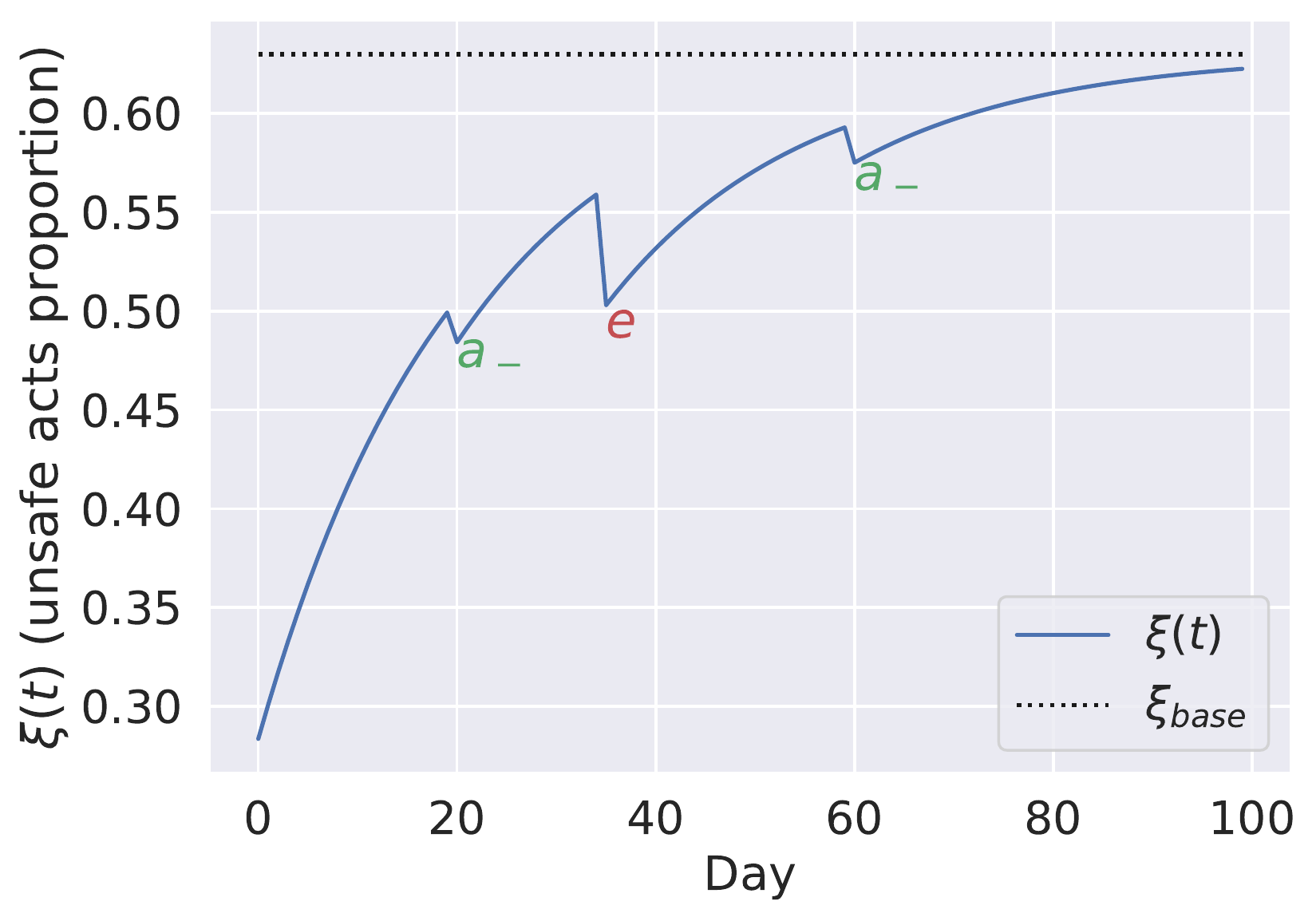}
  \caption{An illustration of the impact of observing unsafe activities/incidents. When unsafe activities \unsafeacts are observed (on day 20 and day 60) it instantaneously drops the proportion of unsafe activities. However, as the time progresses, this proportion returns back to its base value to model the ``forgetting'' aspect in such scenarios. The impact of an incident \incidents (on day 35) is similarly modeled.}
  \label{fig:event_feedback}
\end{figure}

The Intervention process accounts for the dynamic effects on the safety environment due to feedback from incidents and observations. We consider that when an unsafe event \unsafeacts{} or an incident \incidents{} is observed for the $i^{th}$ safety area, immediate feedback is applied to the Event process (indicated by the black dotted arrow in \figref{fig:simulator_schematic}). This feedback could  be due to either an observer's immediate intervention or a post-incident analysis. Accordingly, the feedback is assumed to yield a reduction in the proportion of unsafe activities ($xi^i$), and thus in the number of incidents, in the near future. For instance, if a worker is observed to be welding without the appropriate face-shield, the observer is expected to have promptly pointed out the unsafe activity and suggested corrective action. Therefore one would expect the worker, and perhaps those around it, to be more attentive to safe welding practices in the near future. The impact of such feedback/intervention can be modeled by an instantaneous reduction in the parameter $\xi^i$ after an unsafe activity is observed, as illustrated in \figref{fig:event_feedback}. The impact of observing incidents \incidents{} is similar in every aspect, although more pronounced because a near-miss or an actual incident will usually have a broader and more profound impact on the workforce. Although not implemented here, the impact could be further modulated by the severity of the incident. Also, while  only negative feedback is considered here, the same approach could be similarly applied for a positive observation to account for the benefits from positive reinforcement of safe activities and behaviors~\citep{hollnagel2018safety}.

Mathematically, such feedback (due to intervention) can be incorporated in the time evolution of $\xi^i$ by modifying equation \eqref{eq:theta_dynamics}, as follows. Let $n^i_\unsafeobs$ and $n^i_\incidents$ denote the number of unsafe activities captured by observations of the type $X$ and incidents, respectively, observed at time $t$ for the $i^{th}$ vulnerability. Also, let $\delta_\unsafeobs$ and $\delta_\incidents$ be the user-specified parameters encoding the magnitude of the effect of feedback from an unsafe activity and incident, respectively.  The feedback from these events can be captured by the following dynamic update equation of the safety state,
\begin{align}\label{eq:feedback_effect}
  \theta^i(t)= \theta^i(t-1) + (1-\theta^i(t-1))(n^i_\unsafeobs \delta_\unsafeacts + n^i_\incidents \delta_\incidents).
\end{align}
If no events leading to feedback are observed at time $t$, then $\theta^i(t)$ is updated according to \equationref{eq:theta_dynamics}.

\subsection{Safety Metrics}\label{subsec:health_measures}
Having a simulated environment such as the one described here gives complete access to the true state of the work environment. As shown, the work environment consists of a set of safety areas for which we have defined (in the previous sections) a detailed probabilistic dynamical model. As a result, at each timestep we can assess the safety of the simulated work environment by computing various metrics from the current state of the safety areas. Two such safety metrics are the \emph{daily expected loss} and the \emph{tail-probability} of fatalities. The former quantifies an average loss that is expected to be incurred per day due to safety incidents, whereas the latter corresponds to the probability of  an incident with AHL $\geq$ 4 to occur on any given day. Consider a vector $\mathbf{c}=[0, 1, 10, 100, 1000, 10000]$ expressing the loss values associated with incident severities of Hurt levels 0 through 5, respectively. The units of these values are rather arbitrary but reflect the fact that the impact of an incident increases exponentially with an increase in the incident severity. A similar pattern of exponentially increasing impact with severity can be found in other work (e.g.,~\citet{hallowell2010}). To compute the daily expected loss for the $i^{th}$ safety area, we use the probability distribution of observing $k$ incidents for any given Hurt level; that is, $P(\mathbb{C}^i_{ahl=j}=k)$ for $j \in \{0,1,\cdots, 5\}$, $k \in \mathbb{N}$, as given by \equationref{eq:marginal_dist_ahlj_count}. With this distribution, the daily expected loss for the $i^{th}$ vulnerability can be computed as $\sum_{j=0}^5 \+c_j\cdot\mathbb{E}(\mathbb{C}^i_{ahl=j})$, where the expectation of the incident counts can be derived as follows,
\begin{align}
\mathbb{E}(\mathbb{C}^i_{ahl=j}) &= \sum_{k=0}^\infty k P(\mathbb{C}^i_{ahl=j}=k) \nonumber \\
& = \sum_{k=0}^\infty k  \sum_{t=1}^\infty P(n_\incidents^i=t)P(\mathbb{C}_{ahl=j}^i =k | n_\incidents^i=t), & \text{using \equationref{eq:marginal_dist_ahlj_count}} \nonumber \\
& = \sum_{t=1}^\infty P(n_\incidents^i=t) \sum_{k=0}^\infty k  P(\mathbb{C}_{ahl=j}^i =k | n_\incidents^i=t), & \text{rearranging terms} \nonumber \\
& = \sum_{t=1}^\infty P(n_\incidents^i=t)\cdot t\cdot p^i_j, & \text{using \equationref{eq:HL_count_given_incident_count}}  \nonumber \\
& = \sum_{t=1}^\infty \frac{(\alpha^i \xi^ i \baseRate^i)^t \mathbf{e}^{-\alpha^i \xi^ i \baseRate}}{t!}  \cdot t \cdot p^i_j, & \text{using \equationref{eq:incident_generation}}\nonumber \\
& = \alpha^i \xi^ i \baseRate^i p_j^i. \label{eq:expected_HL_count}
\end{align}

Similarly, one can use the generative model to calculate the tail-probability of high severity incidents with $AHL \ge 4$. For the $i^{th}$ vulnerability this can be obtained using the probability of observing an incident with $AHL=j$, with $j\in \{0,1,\ldots,5\}$, which can be written as
\begin{align}
P^i(AHL=j)= P(n_\incidents^i=0)P(AHL=j|n_\incidents^i=0)  + P(n_\incidents^i\neq0)P(AHL=j|n_\incidents^i\neq0).
\end{align}
For $j>0$, the first term vanishes and the hurt level distribution can be expressed as
\begin{align} \label{eq:HL_distrib}
P^i(AHL=j)= \big(1-\mathrm{Poisson}(0;\lambda^i_* \alpha^i \xi^i)\big)p_j^i,
\end{align}
and the tail-probability can be simply obtained as $\sum_{j=4}^5 P^i(AHL=j)$.
Note that in both equations~\eqref{eq:expected_HL_count} and \eqref{eq:HL_distrib} the time dependence of the parameter $\xi^i$, which is governed by Equations \eqref{eq:xi_dynamics}, \eqref{eq:theta_dynamics} and \eqref{eq:feedback_effect}, is not shown explicitly. Clearly, both of these metrics are calculated at each timestep using the internal state $\xi^i$ to track the performance over time. In the case study shown in the next section, we demonstrate how these safety metrics play a central role in objectively comparing several candidate SA approaches for making safety-related decisions.

\section{Case Study}\label{sec:results}

\begin{table}[tb]
  \begin{center}
  \caption{The parameters values (for each of the safety areas) that fully specify the simulated work environment under consideration. The parameters $\theta(0)$ and $k$ (refer to Equation \eqref{eq:theta_dynamics}) were set to 0.1 and 0.98, respectively, for all safety areas. }
  \label{tab:sim_setting_vuln}
  \vspace*{.6ex}
  \begin{tabular}{c*{9}{l}}
    \toprule
    & \multicolumn{9}{c}{Model Parameters} \\ \cline{2-10}
    Safety Area & $\lambda_*\ $ & $\xi_{base}$ & $\alpha$
      & $p_{0}$ & $p_{1}$ & $p_{2}$ & $p_{3}$ & $p_{4}$ & $p_{5}$ \\
    \midrule
    A   & 17   & 0.55   & 0.04    & 0.50   & 0.35   & 0.13   & 0.02   & 0      & 0      \\ 
    B   & 13   & 0.25   & 0.005   & 0.60   & 0.16   & 0.11   & 0.12   & 0.01   & 0      \\ 
    C   & 22   & 0.4    & 0.01    & 0.30   & 0.06   & 0.35   & 0.28   & 0.01   & 0      \\
    D   & 12   & 0.1    & 0.005   & 0.25   & 0.30   & 0.25   & 0.18   & 0.02   & 0      \\ 
    E   & 15   & 0.05   & 0.01    & 0.38   & 0.26   & 0.16   & 0.16   & 0.03   & 0.01   \\ 
    F   & 5    & 0.45   & 0.02    & 0.58   & 0.06   & 0.08   & 0.18   & 0.08   & 0.02   \\
    G   & 22   & 0.3    & 0.005   & 0.65   & 0.18   & 0.08   & 0.07   & 0.02   & 0      \\
    \bottomrule
  \end{tabular}
  \end{center}
\end{table}

In order to demonstrate the significance of the safety environment simulator for the evaluation of SA approaches, we consider an imaginary work environment with 7 safety areas and 3 observation types. The underlying simulation is described in the previous section and controlled by the parameter values specified in Table~\ref{tab:sim_setting_vuln}. The number of safety areas was kept small for simplicity of presentation. Also, the safety areas and corresponding parameters are not representative of any real work environment. Rather, our aim is to capture the variability of circumstances that SA approaches must deal with in practice. Hence, the simulation parameters were chosen to be diverse and capture distinct aspects of a general work environment, including activities with different rates, propensity to unsafe/at-risk situations, likelihood of safety incidents, and severity of the incidents.

\begin{table}[tb]
  \centering
  \caption{Incident counts for each safety area and every Hurt level over a simulation period of 365~days without feedback. The numbers in table show incident counts at different percentile values from 100 simulations, with the median value (i.e., 50th percentile) outside parenthesis and the p05-p95 range (i.e., 5th to 95th percentiles) inside parenthesis. Recall that AHL=0 indicates near-miss incidents.}
  \label{tab:incident_counts}
  \vspace*{.6ex}
  \begin{tabular}{c*{6}{r@{ }l}}
    \toprule
    & \multicolumn{12}{c}{Incident Counts} \\ \cline{2-13}
    Safety Area
      & \multicolumn{2}{c}{AHL=0} & \multicolumn{2}{c}{AHL=1}
      & \multicolumn{2}{c}{AHL=2} & \multicolumn{2}{c}{AHL=3}
      & \multicolumn{2}{c}{AHL=4} & \multicolumn{2}{c}{AHL=5} \\
    \midrule
    A & 70 & (57--85) & 49 & (38--62) & 19 & (13--27) &  3 & (1--6)   &  0 & (0--0)   &  0 & (0--0)  \\
    B &  4 & (1--6)   &  1 & (0--3)   &  0 & (0--2)   &  1 & (0--2)   &  0 & (0--0)   &  0 & (0--0)  \\
    C & 10 & (5--16)  &  2 & (0--4)   & 11 & (6--17)  &  9 & (5--13)  &  0 & (0--1)   &  0 & (0--0)  \\
    D &  0 & (0--2)   &  1 & (0--2)   &  0 & (0--2)   &  0 & (0--2)   &  0 & (0--0)   &  0 & (0--0)  \\
    E &  1 & (0--3)   &  1 & (0--2)   &  0 & (0--2)   &  0 & (0--2)   &  0 & (0--1)   &  0 & (0--0)  \\
    F & 11 & (7--17)  &  1 & (0--3)   &  2 & (0--3)   &  3 & (0--7)   &  1 & (0--3)   &  0 & (0--2)  \\
    G &  6 & (3--11)  &  1 & (0--4)   &  1 & (0--2)   &  1 & (0--2)   &  0 & (0--1)   &  0 & (0--0)  \\
    \bottomrule
  \end{tabular}
\end{table}

For reference, Table~\ref{tab:incident_counts} shows the incidents counts for each of the safety areas and Hurt levels over the duration of a simulation if no intervention is applied. Since the simulation can be repeated multiple times, one can obtain statistics over the incident counts, such as the median value (i.e., 50th percentile) and the p05-p95 range (i.e., 5th to 95th percentiles) of the number of incidents, as shown in the table for each safety area and Hurt level. The ranges of outcomes in Table~\ref{tab:incident_counts} demonstrate the encoded variability in the simulated environment. Moreover, one can also  observe the correspondence between the overall number of incidents and the simulation parameters specified in Table~\ref{tab:sim_setting_vuln} for the different safety areas. Furthermore, the distribution of incident counts between safety areas and Hurt levels illustrates the diversity in inherent risks associated with different safety areas. For instance, safety area~A was specified with high task rate ($\lambda_*=17$) but a tendency toward low-severity incidents. This safety area could correspond to common activities (e.g., walking on site) or frequent safety aspects (e.g., \emph{hand safety}) that have very low chance for high-severity incidents. In contrast, safety area~F illustrates activities performed much less frequently ($\lambda_*=5$) but with a significant chance of severe consequences in the event of an incident. This safety area could correspond, for example, to \emph{crane lifting}/\emph{moving heavy equipment} work.

\begin{table}
  \centering
  \caption{Simulation parameters with respect to each observation type. The parameters $m$ and $\delta_{\unsafeacts}$ denote the number of observations to be made per day and the relative reduction in incident rate of a safety area when a negative observation is recorded, respectively. The parameters $\eta_{\unsafeacts}$ and $\eta_{\safeacts}$ model any potential biases in the observation process. In the current setting, WSO and BPO observation types are biased (the former is biased towards recording more negative observations and the latter vice versa), while SAO is unbiased.}
  \label{tab:sim_setting_obs}
  \vspace*{.6ex}
  \begin{tabular}{ccccc}
    \toprule
    Observation Types & $m$ & $\delta_{\unsafeacts}$ & $\eta_{\unsafeacts}$ & $\eta_{\safeacts}$ \\
    \midrule
    WSO & 2 & 0.03 & 100 & 150 \\
    SAO & 2 & 0.03 & 100 & 100 \\
    BPO & 1 & 0.03 & 120 & 100 \\
    \bottomrule
  \end{tabular}
\end{table}

As mentioned at the onset, the main motivation for the simulator was to provide a platform to test various SA approaches and evaluate their ability to help identify and mitigate higher-risk safety areas. Here we consider a particular case where SA approaches determine the allocation of limited observation resources to different safety areas based on their perceived risk. Considering that observations provide an opportunity to assess the overall safety and correct unsafe situations, the optimal allocation of observation resources is essential. Moreover, by optimally assigning observation resources, one can improve the safety of the work environment by providing constructive feedback where it is needed the most. In other words, the allocation should ideally reflect the safety risk associated with each safety area. Consider that we have finite resources to make the three types of observations: WSO, BPO and SAO in out case (refer to \secref{sec:observations} for details). In the simulator these observations are characterized by the parameters listed in Table~\ref{tab:sim_setting_obs}, as detailed in \secref{subsec:observation_process}. Then, as discussed in \secref{subsec:intervention_process}, unsafe/at-risk observations of the $i^{th}$ safety area have an impact in its state $\theta^i(t)$ with $\delta_\unsafeacts=0.03$ for all observation types. For the purpose of this case study, we ignored incident feedback ($\delta_\incidents=0.0$); a choice that ensured that the assessment is a direct reflection of a SA approach's ability to correctly identify underlying risks and allocate resources accordingly. 

We consider four simple approaches, the first being a \emph{random} approach allocating the 5 observers (cf.~Table~\ref{tab:sim_setting_obs}) randomly across 7 safety areas on any given day. This approach is clearly suboptimal unless all the safety areas have exactly the same risks. An alternative approach is to allocate observers on safety areas according to their relative \emph{number of incidents} in the past 30~days. This approach seems reasonable because more incidents is an indication that a safety area is likely to have an higher number of unsafe activities and thus is more likely to experience future  incidents. On the other hand, such an approach does not consider the severity of the incidents.
Hence, yet another approach (third) would be to allocate the observers based on the highest \emph{incident severity} in the past 30~days. For instance, one could assign to the $i^{th}$ safety area a weight that increases exponentially with the incident severity. For this example, we assign a weight of $2^{h^i(t)}$, where $h^i(t)$ is the highest incident severity in the last 30~days, and then allocate observers by normalizing these weights over safety areas (i.e., such that they sum to one). The fourth approach considers a \emph{weighted random} allocation of observers to focus the observation resources more toward certain safety areas than others, based on predefined weights. For instance, due to a good foresight it was anticipated that safety area~F would have a significantly higher propensity to encounter high-severity incidents, while the safety areas~D and E are expected to produce a very low number of incidents because of strict controls in place. Based on this information, one could decide to allocate resources randomly with the following proportionality weights: [\vulnweight{A}{0.12}, \vulnweight{B}{0.12}, \vulnweight{C}{0.12}, \vulnweight{D}{0.08}, \vulnweight{E}{0.08}, \vulnweight{F}{0.28}, \vulnweight{G}{0.2}]. Based on these weights, over a course of several days safety area~F will get more than a quarter of available resources and safety areas~D and E less than a tenth. This approach represents here a ``best-case scenario'' because it reflects the case in which a~priori information of both the underlying frequency of incidents and their severity distribution is known.  Clearly, the above approaches are very simple and only two are informed by data. This is done deliberately as including an advanced, more complex SA approach may needlessly confound the analysis.

\begin{figure}
  \centering
  \includegraphics[width=\textwidth]{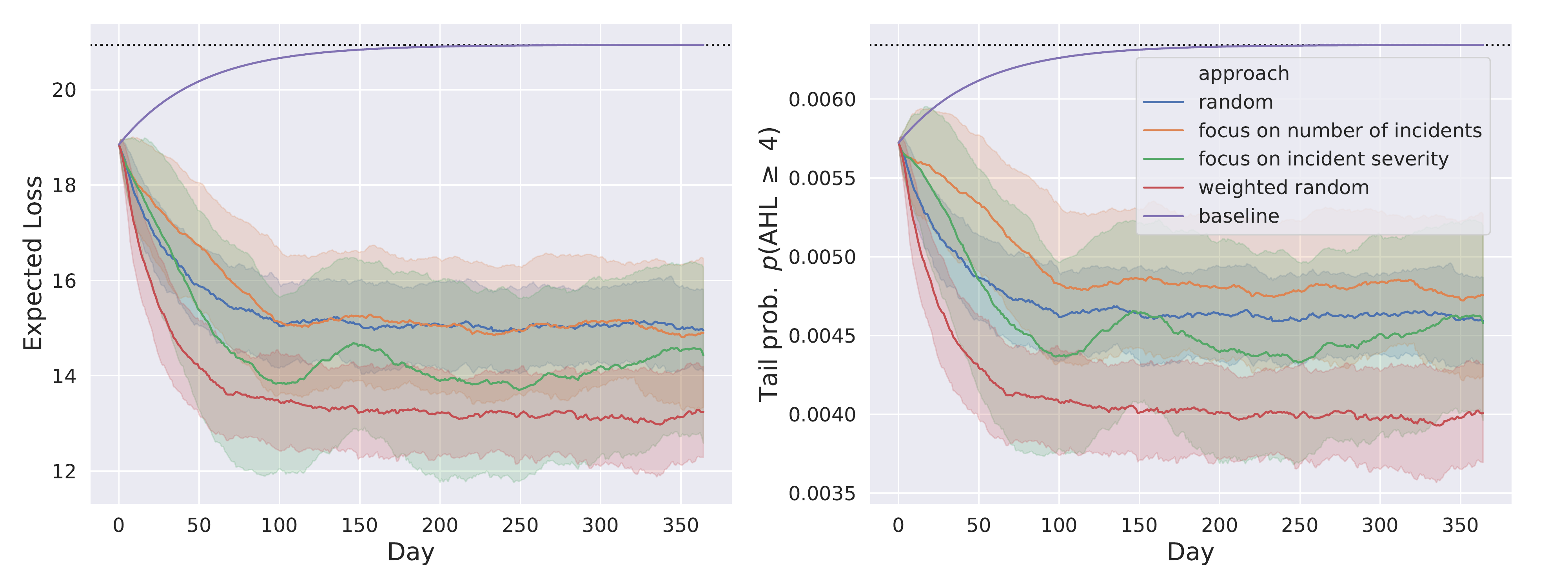}
  \caption{Comparison of SA approaches for allocating observation resources using the safety metrics \emph{expected loss} and \emph{tail probability} (cf.~\secref{subsec:health_measures}). The baseline corresponds to the no-observation case, meaning that no observers are allocated for conducting inspections, and thus the environment's state deteriorates at a natural rate as governed by the simulator's dynamics to the its limiting value indicated by the dotted line. The plots are obtained by performing 100 simulation runs with the same initial state. The lineplots show the mean value and $\pm$1 standard deviation uncertainty band. Note that the baseline curve is deterministic since there is no feedback (no observations).}
  \label{fig:random_policies_comparison}
\end{figure}

Since each simulation is stochastic in nature, one needs to run the simulator several times in order to test the SA approaches.
In this case, results were obtained by running 100 simulations with daily timesteps over a total of 365~days. During each simulation the safety metrics, discussed in \secref{subsec:health_measures}, were recorded on daily basis and shown in \figref{fig:random_policies_comparison}. As a reference, baseline curves are also plotted, corresponding to the case when no observations are conducted to monitor and address unsafe activities that may be present. The baseline curve is the absolute worst case scenario because, without any monitoring and feedback, the safety state of the environment simply deteriorates at the rate specified by $k^i$ (cf. \equationref{eq:theta_dynamics}) toward the worst-case base level, indicated by the dotted line. Note that in the baseline scenario the two safety metrics evolve deterministically because in absence of observations the environment gets no feedback. On the other hand, when observations are made (based on any SA approach), the resulting feedback from interventions changes the state of the environment, thus leading to the observed stochasticity in the two safety metrics.

The results in \figref{fig:random_policies_comparison} distinctly contrast the different approaches with respect to the simulated  environment. Remarkably, under these settings, the approach allocating observer according to the number of incidents performs worse than the random approach in terms of tail-probability and comparably in terms of expected loss. While the former approach does reduce the overall number of incidents, it does not take the severity of incidents into account and is thus biased toward safety areas with many incidents even if those have lower impact. The approach based on the severity of incidents faired better in both metrics because, by design, it tries to focus on safety areas with higher severity incidents. The plots also clearly establish, unsurprisingly, that the weighted random approach performs the best in this case because, as mentioned, it uses a correct prior information about the risk of different safety areas.
The reader is cautioned not to draw any specific conclusion about the effectiveness of a SA approach over another, as the main point of this analysis was to simply demonstrate how the simulator enables systematic testing of various SA approaches. 
\begin{table}
  \centering
  \caption{Incident counts per severity at the end of \emph{one simulation} for the safety approaches considered in the case study.}
  \label{tab:onerun.incident.counts}
  \vspace*{.6ex}
  \begin{tabular}{lcccccc}
    \toprule
    & \multicolumn{6}{c}{Incident Counts} \\ \cline{2-7}
    Approach & AHL=0 & AHL=1 & AHL=2 & AHL=3 & AHL=4 & AHL=5 \\
    \midrule
    baseline                     &  103 &  52 &  40 &  15 &  2 &  0 \\
    random                       &   72 &  26 &  27 &  18 &  2 &  0 \\
    focus on number of incidents &   48 &  21 &  10 &  10 &  1 &  0 \\
    focus on incident severity   &   58 &  29 &  20 &  13 &  1 &  0 \\
    random weighted              &   69 &  40 &  25 &   8 &  1 &  0 \\
    \bottomrule
  \end{tabular}
\end{table}

It is also worth reiterating the limitations of testing a safety approach on only one data realization as previously mentioned. To  demonstrate this, consider the incident counts per severity after only one simulation run shown in Table~\ref{tab:onerun.incident.counts}. From these results, one would come away with a markedly different conclusion that responding to the number of incidents was the most effective approach since it yielded the lowest number of incidents across nearly all severity levels. While that is true in this realization, drawing such a conclusion fails to consider the large negative impact when the rare, high-severity incidents do occur. Hence, although focusing on reducing the number of incidents indeed yields a lower overall number of incidents, that is likely not the most effective approach to reduce high-severity incidents in a long run.


\section{Conclusion and discussion}\label{sec:discussion}


With increasing work on data-driven approaches to improve safety performance, it has become imperative to reconsider the techniques used to test and evaluate them. 
To that end, this article presents a methodology wherein any safety analytics (SA) approach can be tested within a simulated industrial work environment and its afety management system (SMS). The simulator was designed such that it incorporated a number of general characteristics that were elicited from safety experts. The simulator models these characteristics
through the specification of three stochastic processes viz. \emph{event process}, \emph{observation process} and \emph{intervention process}, as illustrated in \Figref{fig:simulator_schematic} and detailed in \secref{sec:simulator}.  
One potential concern with the decomposition of the methodology into three interdependent processes is  that to what extent such separation exists in real-world. To that we reiterate that the simulator is not intended as a perfect model of reality, but rather as an emulator capable to generating safety data so as to objectively test SA approaches under conditions that mimic, albeit roughly, our understanding of industrial work environments.

Mathematically, the simulator is specified as a collection of probabilistic models~\citep{gelman2014bayesian, koller2009probabilistic}. This means that all of the different elements leading to the safety data are simulated as realizations from  interdependent random processes, which encode natural couplings between different factors such as task rates, fraction of unsafe/at-risk activities, frequency of incidents with different severities etc. We note that the simulator makes certain simplifying assumptions, but that does not compromise its ability to test and compare SA approaches. Nonetheless, a number of alternate choices (to the ones we have made) are possible. For example, one could have use a renewal point process (instead of a Poisson process) to characterize the distribution of events over time, which introduces dependencies between events~\citep{daley1988introduction}, or considered other ways to model the human tendency of forgetfulness over time, or another mechanism to account for a safety intervention/feedback.

The simulator's processes are controlled by several user-selected parameters (see Table~\ref{tab:sim_setting_vuln} for those used in the case study example). These parameters have a simple interpretation in terms of what role they play in reflecting different aspects of real work environments and SMSs. These are considered input parameters because their selection depends primarily on which characteristics one wants to test a SA approach on. Of course, one can estimate values and trends from historical data so that the simulator mimics a particular work environment. It should be noted that one can also choose these parameters beyond those anticipated to be encountered in practice. For instance, one could test the robustness of a SA approach in worst case scenarios in which the safety data is either sparse or has large amounts of bias or both. The ability to specify these simulator parameters is crucial to obtain a controllable and repeatable testbed for comparing SA approaches.

Since the methodology proposed here is based on a simulation of reality, one might question how significant are the results. This is an understandable concern, but it must be emphasized that this is, to the best of our knowledge, the only way to systematically test SA approaches that addresses many challenges with SMSs. Moreover, the presentation in \secref{sec:simulator} attempted to show how the construction of the simulator closely mirrors the best understanding of industrial work environments from safety experts. It is expected that the performance of SA approaches will be dependent on the specific values chosen for the simulator parameters. However, having the ability to modify those parameters allows for enough flexibility to create a representative and meaningful testbed which can then pave the way for a fair and systematic comparison of different SA approaches, as demonstrated in \secref{sec:results}.

A feature that is missing in the proposed simulator is the generation of any contextual information in the form of a narrative (text description) associated with a safety event. The simulator, in its present form, is limited to capturing the relationships and statistics of the different events as these are arguably the most common summary elements considered by decision makers. However, adding the capability to generate textual description of safety events is conceptually straightforward. Doing so will invariably result in a significant increase in complexity of the simulator, but it would primarily be driven by the complexity in generating realistic text. The addition of this aspect could be a potential avenue for future work. Similarly, another avenue that could be beneficial to the advancement of safety science is to investigate additional safety metrics (other than the proposed metrics of \emph{expected loss} and \emph{tail-probability}) that can further facilitate meaningful evaluation of different SA approaches.

Lastly, while the motivation for this article was on testing of SA approaches, it must be noted that the methodology could be used to evaluate potential uplift of nearly any change to an existing SMS. Since the simulated environment provides the means to compare two (or more) SA approaches, the methodology can be used as a general testbed for new ideas, concepts, proposed changes, and even investigation around which additional types of safety data should be collected. Of course, depending on what is being tested, doing this might require the simulator to be extended such that it incorporates the additional elements under consideration. However, as previously mentioned, the complexity of generating a specific data modality might be methodologically substantial but is conceptually straightforward within the proposed framework. 


 \section{Acknowledgements}
 The authors would like to thank Nathaniel Rogalskyj for helpful discussions in the development of the work safety environment simulator, and ExxonMobil's management for their continued support to advance the critical area of industrial safety risk management.

\bibliographystyle{elsarticle-harv}
\bibliography{refs}

\end{document}